\documentclass[reprint,%
aps,prd,%
10pt,%
superscriptaddress,%
amsmath,amssymb,%
floatfix]{revtex4-2}

\usepackage{amsmath,amssymb,bm}
\usepackage{mathtools}
\usepackage{esvect}
\usepackage{booktabs}
\usepackage{multirow}
\usepackage{array}
\usepackage[svgnames]{xcolor}
\usepackage[colorlinks=true,%
linkcolor=MediumBlue,%
urlcolor=MediumBlue,%
citecolor=MediumBlue]{hyperref}
\renewcommand{\vec}[1]{\ensuremath{\mathbf{#1}}}

\newcommand{\Like}{\mathcal{L}}

\begin{document}

\title{Search for non-Newtonian interactions at micrometer scale with a levitated test mass}

\author{Charles P. Blakemore}
\affiliation{Department of Physics, Stanford University, Stanford, California 94305, USA}

\author{Alexander Fieguth}
\affiliation{Department of Physics, Stanford University, Stanford, California 94305, USA}

\author{Akio Kawasaki}
\thanks{Now at National Metrology Institute of Japan (NMIJ), National Institute of Advanced Industrial Science and Technology (AIST), 1-1-1 Umezono, Tsukuba, Ibaraki 305-8563, Japan}
\email{akio.kawasaki@aist.go.jp}
\affiliation{Department of Physics, Stanford University, Stanford, California 94305, USA}
\affiliation{W. W. Hansen Experimental Physics Laboratory, Stanford University, Stanford, California 94305, USA\looseness=-1}

\author{Nadav Priel}
\email{nadavp@stanford.edu}
\affiliation{Department of Physics, Stanford University, Stanford, California 94305, USA}

\author{Denzal Martin}
\affiliation{Department of Physics, Stanford University, Stanford, California 94305, USA}

\author{Alexander D. Rider}
\thanks{Now at SRI International, Boulder, Colorado 80302}
\affiliation{Department of Physics, Stanford University, Stanford, California 94305, USA}

\author{Qidong~Wang}
\affiliation{Institute of Microelectronics of the Chinese Academy of Sciences, Beijing 100029, China}

\author{Giorgio Gratta}
\affiliation{Department of Physics, Stanford University, Stanford, California 94305, USA}
\affiliation{W. W. Hansen Experimental Physics Laboratory, Stanford University, Stanford, California 94305, USA\looseness=-1}

\date{\today}

\begin{abstract}

We report on a search for non-Newtonian forces that couple to mass, with a characteristic scale of ${\sim}10~\mu$m, using an optically levitated microsphere as a precision force sensor. A silica microsphere trapped in an upward-propagating, single-beam, optical tweezer is utilized to probe for interactions sourced from a nanofabricated attractor mass with a density modulation brought into close proximity to the microsphere and driven along the axis of periodic density in order to excite an oscillating response. We obtain a force sensitivity of ${\lesssim}10^{-16}~\rm{N}/\sqrt{\rm{Hz}}$. Separately searching for attractive and repulsive forces results in the constraint on a new Yukawa interaction of $|\alpha| \gtrsim 10^8$ for $\lambda > 10~\mu$m. This is the first test of the inverse-square law using an optically levitated test mass of dimensions comparable to $\lambda$, a complementary method subject to a different set of systematic effects compared to more established techniques.

\end{abstract}

\maketitle

\section{Introduction}

Among fundamental interactions, gravity has the distinction of simultaneously being the most apparent and yet the least understood. From the theoretical point of view, the universal law of gravitation~\cite{Newton} and general relativity~\cite{Einstein} have been successful in describing interactions at macroscopic scale. However, unlike other fundamental interactions such as electromagnetism, empirical knowledge of gravity at sub-millimeter scale is rather rudimentary. At the same time, connections between gravitation and quantum mechanics are still obscure, yet much of theoretical physics has been driven by the assumption that gravity remains unmodified all the way down to the Planck scale.
Modifications of gravity in such a large and poorly constrained region of parameter space could guide us toward solutions of outstanding theoretical quandaries such as the hierarchy problem, the dark matter puzzle, and the unification of gravity with the Standard Model of particle physics~\cite{ArkaniHamed:1998rs,Antoniadis:1998,Adelberger:2003,Mukohyama:2016,Sundrum:2004,Nelson:2011,Graham:2016,Adelberger:2009}.

It is customary to modify the inverse square law (ISL) of Newtonian gravity by introducing an additional Yukawa potential with a length scale $\lambda$. The resulting potential between two point masses can be written as:
\begin{equation}\label{eq:NNG}
    V(r) = - G_{\infty} \frac{M_1 M_2}{r} (1 + \alpha e^{-r/\lambda}),
\end{equation}
\noindent with $G_{\infty}$ the Newtonian constant of gravitation, $M_1$ and $M_2$ the gravitating masses, $r$ their distance, and $\alpha$ the relative magnitude of the new interaction. $\alpha$ can be either positive or negative, and may depend on properties such as mass or baryon number~\cite{Adelberger:2003}. 

Traditionally, gravitational interactions have been experimentally investigated using sophisticated torsion balances~\cite{Cavendish} which establish some of the most stringent bounds on deviation from the ISL at sub-millimeter scale~\cite{Adelberger:2003,Hoyle:2001,Hoyle:2004,Kanper:2007,Lee:2020,Tan:2020}. Alternative techniques have been developed using nanotechnology to mount test masses at the ends of microcantilevers~\cite{Geraci:2008,Chen:2016,Sushkov:2011}. Generally, all measurements within this field are limited by systematic effects, such as the reliability and reproducibility (or lack thereof) in the positioning and alignment of the macroscopic objects involved, especially given the small separations required for competitive measurements. Hence, experimental progress calls for new techniques with different attributes and systematics that may eventually contribute to robust discoveries.

In the present work, we describe the first investigation of the ISL in the $1<\lambda<100~\mu$m range using an optical tweezer in vacuum, where radiation pressure is used to counter the Earth's gravity and to provide the restoring force against which the interaction is compared. As first discussed in~\cite{Geraci:2010b}, the motion of an optically levitated silica microsphere~\cite{Ashkin:1970,Ashkin:1971} (MS) is studied to infer its coupling with an attractor system (AS) in which regions of different mass density 
are alternated on a microscopic scale. To our knowledge, this is the search using the smallest objects to both source and sense a new interaction or modified gravity.  So far, experiments probing the micrometer regime have been mainly conducted with greater separations between the source and the test mass, and/or using substantially larger test and source masses. In this study, the separation between MS and AS, the scale of the test mass, and the AS density modulation, are all matched to the length scale $\lambda$ of the interaction. This results in measurements with broader applicability, including to non-Newtonian potentials that cannot be described by the form in Eq.~(\ref{eq:NNG}).


The MS, acting as a force sensor, is isolated from the environment so that its center of mass motion can be reduced to very low effective  temperatures~\cite{Delic:2020} in an otherwise room temperature setup. The charge state of the MS can be controlled with exceptional accuracy~\cite{Moore:2014} to provide an empirical force calibration and, during ISL test measurements, ensure overall neutrality. Directly measuring the force vector on the MS~\cite{Blakemore:2019} provides more dimensions to understand backgrounds and provides sensitivity to the sign of $\alpha$, in contrast to experiments only sensitive to a deviation from $|\alpha|=0$~\cite{Chen:2016,Sushkov:2011,Geraci:2008}. Finally, many methods developed in quantum optics can be applied to this technique in the future, with the potential for substantial advances in an all important problem of experimental physics.  

\begin{figure}[!t]
    \includegraphics[width=1\columnwidth]{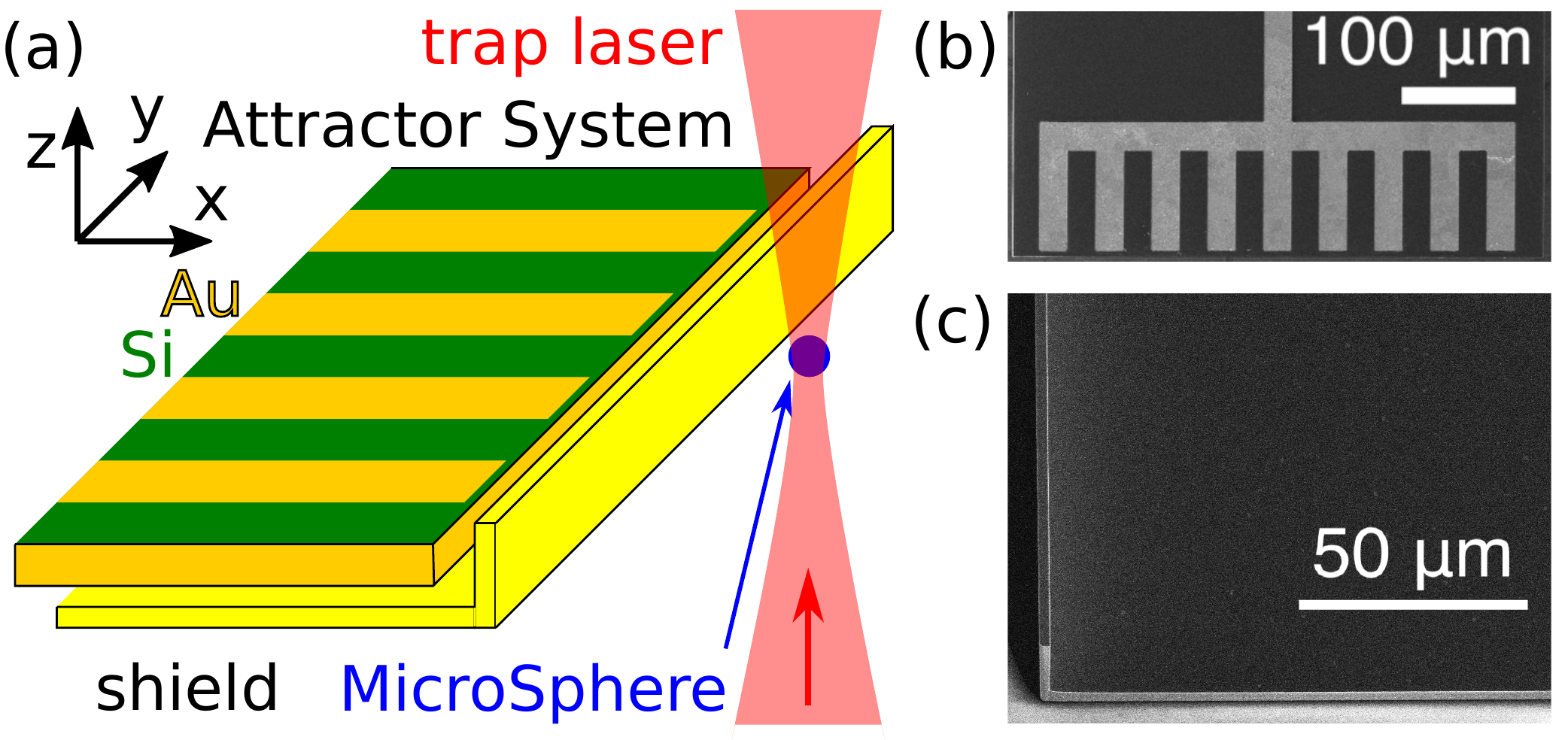}
    \caption{(a) Central portion of the experimental setup: a MS is trapped in an optical tweezer. A stationary shield centered about the trapped MS, with the closest surface within a few microns of the MS, and the AS is behind it. (b) Scanning electron microscope (SEM) image of the AS. The dark (bright) regions correspond to silicon (gold). (c) SEM image of the shield, viewed at a 40$^{\circ}$ angle to highlight the three-dimensional structure, with the vertical wall to the left.}
    \label{fig:expsetup}
\end{figure}

\section{The experiment}

The overall apparatus layout, MS trapping, force calibration, charge neutralization, metrology, and the force sensitivity achieved, are described in detail in Ref.~\cite{Kawasaki:2020}. Briefly, the central part of the system, shown in Fig.~\ref{fig:expsetup}, is a $7.56\pm0.19~\mu$m diameter silica MS~\cite{microparticles_gmbh} trapped in an upward-propagating, single-beam optical tweezer, formed by 1064~nm light focused down to a waist size of $3.2~\mu$m by an off-axis parabolic mirror with a focal length of $5~$cm. The mass and density of the MS are estimated to be $m=414\pm15~$pg and $\rho = 1.83\pm0.15~\rm{g/cm^{3}}$ from a combination of measurements \emph{in situ} for another MS from the same batch, following the method in Ref.~\cite{Blakemore:2019_2}, and manufacturer's specifications~\cite{microparticles_gmbh}.

The $x$ and $y$ positions of the MS are measured by interfering the recollimated forward-scattered light with a reference wavefront and projecting the result onto a quadrant photodiode (QPD). The $z$ position of the MS is measured by interfering the light retroreflected by the MS with another reference wavefront, whereby motion along $z$ produces a change in the path length and thus in the phase of the retroreflected light. Both interference measurements make use of heterodyne detection, in which the reference wavefronts are frequency-shifted by $-125~$kHz relative to the trapping beam. The photocurrent signals are then amplified, digitized, and digitally demodulated. The resulting measurements of the $x$, $y$, and $z$ degrees of freedom are used both for real-time feedback control and offline analysis.

The trapping region is surrounded by six identical electrodes resulting in a cubic cavity in which the MS is shielded from external electric fields. The electrodes have holes for optical and mechanical access from six directions, and they can be individually biased to control translational and rotational degrees of freedom of the MS.  This feature is used to calibrate the force sensitivity of the system by adding a well-defined charge to the MS
and driving its motion with AC fields applied to the three pairs of opposite electrodes~\cite{Moore:2014,Rider:2016,Rider:2018,Blakemore:2019,Kawasaki:2020}. These manipulations are generally done with the AS and shield in their retracted position, so that the applied electric field at the MS location is well understood and approximately uniform. 

Prior to the ISL measurements, the neutral MS is driven to rotate at 6~kHz, by coupling a rotating electric field to the permanent electric dipole moment in the MS~\cite{Rider:2019,Blakemore:2020}. This results in a lower and more consistent force noise. At the $4\times 10^{-7}~$mbar vacuum employed here, the MS's angular velocity decays exponentially with a time constant ${>}8$~hours~\cite{Blakemore:2020} in the absence of a driving field and while ISL measurements are performed. The natural oscillation frequency of the trapped MS is ${\sim}380~$Hz for both $x$ and $y$, while feedback in the $z$ direction results in a similar trapping frequency (cf. the optical spring constant without the feedback in the $z$ direction corresponds to ${\sim}30~$Hz). Slow drifts in the $z$ position, which may be attributed to changes in the optical path, are corrected at ${\sim}10~$s intervals by an auxiliary measurement performed using a camera-based microscope installed at a side-view port.

The AS (Fig.~\ref{fig:expsetup}b) is a cantilever device, nanofabricated in silicon and measuring $500~\mu$m$~\times~475~\mu$m$~\times~9~\mu$m in the $x$, $y$, $z$ directions, and supported by a thick silicon handle~\cite{Wang:2017}. The front portion of the AS, closest to the trapped MS, is patterned with nine rectangular trenches filled with gold, regularly spaced along the $y$ axis with a pitch of $50~\mu$m, measuring $25~(100)~\mu$m in the $y$ ($x$) direction to create the required density modulation. 

Although the AS is coated with $150~$nm of gold over a $50~$nm titanium adhesion layer, a separate shield is employed to further reduce both scattered light and electrostatic backgrounds.  The shield (Fig.~\ref{fig:expsetup}c) is also nanofabricated in silicon, to obtain an L-shaped cross-section in the $x-z$ plane. The horizontal plane of this device is $350~\mu$m$~\times~1000~\mu$m$~\times~3~\mu$m in the $x$, $y$, $z$ directions, and the vertical wall nearest to the trap is $22~\mu$m tall ($z$) and ${\sim}2~\mu$m thick ($x$). The shield, also sputter-coated with $150~$nm gold over $50~$nm titanium, is maintained stationary during a measurement, while the AS scans along the $y$ direction with reciprocating motion. This arrangement is designed to reduce the background from electric field gradients, originating from both a contact potential and patch potentials of the surface of the AS~\cite{Garret:2015,Blakemore:2019}, as it scans in front of the MS. Additionally, the shield reduces backgrounds due to modulations of the halo of the trapping beam or other stray light, which mimic minute shifts in the centroid of light on the QPD. 

With all devices in position as in Fig.~\ref{fig:expsetup} and the apparatus calibrated as described, the AS undergoes harmonic reciprocating motion with a frequency of $3~$Hz and a peak-to-peak amplitude of $202~\mu$m along the $y$ direction, corresponding to ${\sim}4$ full periods of the density modulation. During a 10-s-long measurement, the motion of the MS, the position of the AS in three dimensions, as well as various power-monitoring photodiodes, and feedback monitors, are synchronously digitized at $5~$kHz and stored in a single binary file with timestamps.  Environmental variables such as temperature and atmospheric pressure are sampled at a lower rate stored separately. A total integration of $10^5~$s is obtained by repeating such $10~$s measurements $10^4$ times.

\section{Analysis}
\subsection{The data sample}

\begin{figure}[t]
    \includegraphics[width=1\columnwidth]{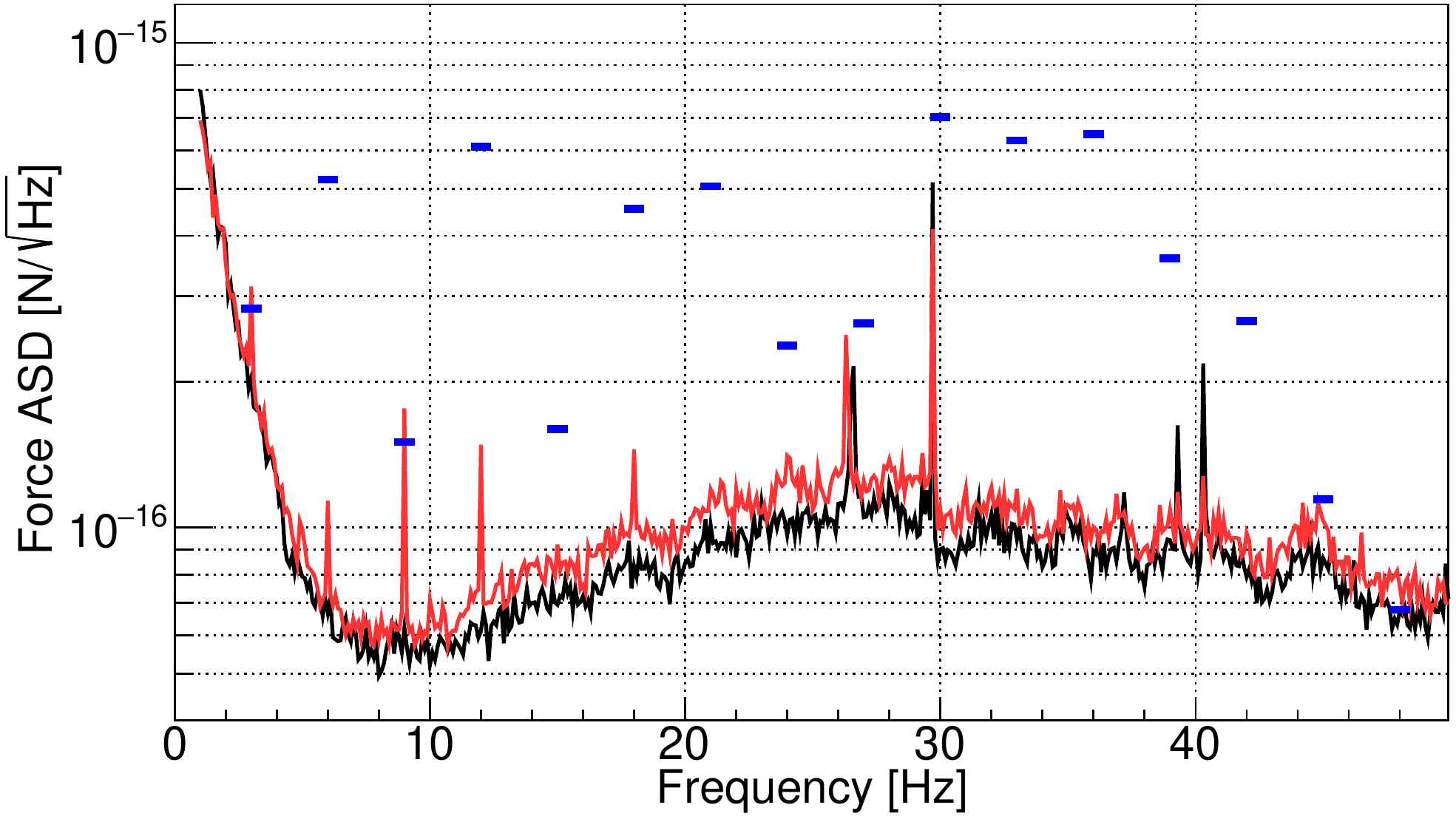}
    \caption{Amplitude spectral density (ASD) of the $z$ component of the force on a $7.56~\mu$m diameter MS. The black (red) curve shows actual data with the AS stationary (scanning along $y$ at $3~$Hz with $202~\mu$m peak-to-peak amplitude). The blue bars show a comparison to an expected MS response produced by the potential described by Eq.~\ref{eq:NNG} with with $\alpha=10^{10}$ and $\lambda=10~\mu$m. The data displayed here is the average of 100 distinct $10$-s integrations.}
    \label{fig:asd}
\end{figure}

For the $7.56~\mu$m silica MSs used here, a force sensitivity of ${\leq}1\times10^{-16}~{\rm N}/\sqrt{\rm Hz}$ in the $1~$Hz to ${\sim}50$~Hz frequency range is achieved~\cite{Kawasaki:2020}. For neutral MSs, this performance is also observed when both AS and shield are in close proximity, as shown by a typical force amplitude spectral density (ASD) displayed in Fig.~\ref{fig:asd}, with the closest shield surface at $11~\mu$m from the center of the MS.  The observed baseline noise is of a statistical nature, and can be integrated for multiple days without encountering an irreducible floor. 
The $10^5~$s data set used here was collected with one MS. The distance between the center of the MS and the front surface of the AS in the $x$ direction is $13.9~\mu$m, and the offset between the center of the MS and the center of the AS is $4.9~(-15.7)~\mu$m in the $y$ ($z$) direction. The uncertainties and drifts of these parameters over the entire run are about ${\pm}1~\mu$m or less and are specifically shown in Table~\ref{tab:sysuncert}.  Although the expected sensitivity for this exposure at the noise limit corresponds to $\alpha\approx 1\times10^7$ for $\lambda=10~\mu$m, the actual sensitivity is limited by backgrounds, which manifest when the AS scans. This is illustrated by Fig.~\ref{fig:asd}, as there are specific frequencies at which a response well above the noise results from the scanning of the AS. 

\subsection{Backgrounds}

Backgrounds can originate from several sources. Interactions between electric field gradients induced by the AS and the electric dipole moment of the MS, estimated to be $10^2-10^3~e{\cdot}\mu$m~\cite{Rider:2016,Rider:2019,Blakemore:2020} with $e$ the fundamental charge, are expected in all directions, with different levels of attenuation from the shield.  In the $xy$-plane, backgrounds may also arise from small variations in the halo or stray light, driven by the scanning motion of the AS. In the $z$ direction, this background is expected to be substantially smaller as the shield blocks the AS in the image plane of the retroreflected photodiode, although couplings between $z$ and $x$-$y$ at the $20\%$ level exist. The $x$-$y$ components of the background observed at individual frequencies are as large as $1.5 \times 10^{-15}~$N, which is equivalent to $\alpha \gtrsim 10^{11}$ for $\lambda=10~\mu$m. 

While the three dimensions can eventually be used to provide a more sensitive measurement, the asymmetry in the current background levels makes the measurement along $z$ substantially superior for the present analysis.  By modeling the system with a finite element method, it was found that a contact potential difference of ${\sim}50~$mV between the AS and the shield can account for backgrounds in $z$ at the observed order of magnitude.  Backgrounds from patch potentials on the AS are found to be subdominant because of strong attenuation from the shield.

\subsection{Signal model}
In order to conduct a search for non-Newtonian forces that couple to mass, a signal model is built from mesh calculations of the force between the AS and MS as a function of their relative displacement, for various length scales $\lambda$. The signal scales proportionally to $\alpha$, which is the parameter of interest in the statistical inference procedure. The model is sampled by the measured position of the AS during each $10~$s run to generate the expected force on the MS as a function of time. The MS response is expected to have different amplitudes at several integer multiples of the fundamental frequency $f_0$ of the AS motion, as shown Fig.~\ref{fig:asd}.  

As some background sources, such as vibration, are expected to affect mainly the fundamental frequency, we exclude $3~$Hz and use only harmonics which contain an expected signal stronger than that of $3~$Hz. Also excluded are the 6~Hz, $2^{nd}$ harmonic, because of a potential background arising from nonlinearities in the system, and the $30~$Hz, $10^{th}$ harmonic, because of an unidentified large spectral feature at $29.7~$Hz (also present with the AS stationary).  Therefore, the search is performed using the harmonics at $12$, $18$, $21$, $33$, $36$, and $39~$Hz. In addition to the amplitude information, the phase of the expected signal relative to the AS motion is incorporated for all those harmonics.  

\begin{figure}[!t]
    \includegraphics[width=1\columnwidth]{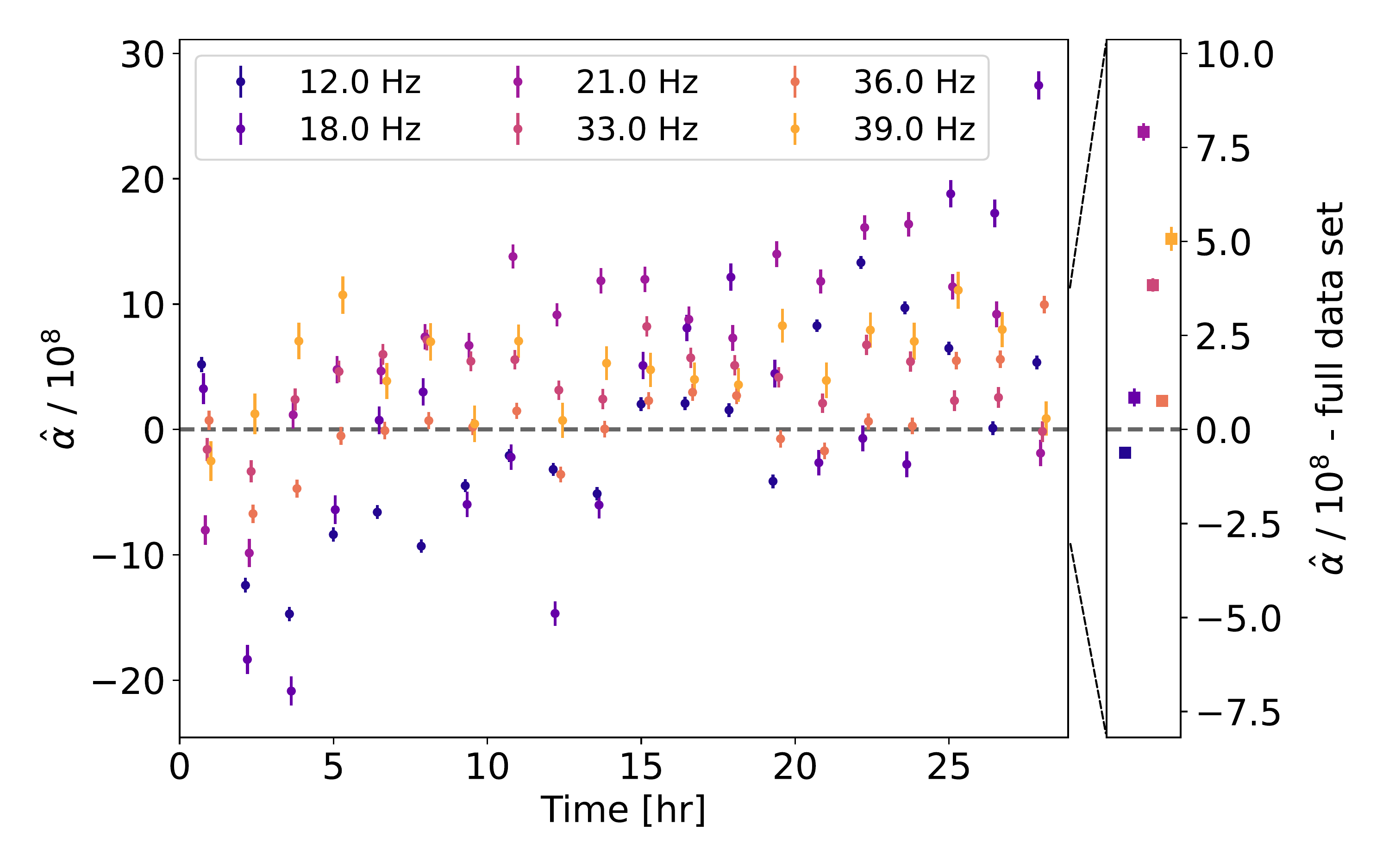}
    \caption{The single harmonic maximum likelihood estimator (MLE) $\hat{\alpha}_i$ for $\lambda=10~\mu$m as a function of time for the six harmonics used in the analysis. Each harmonic $f_i$ is evaluated separately taking into account its own phase response and noise level. Here, each estimation of $\hat{\alpha}_i$ comes from 5000~seconds of data. The error bars represent 95\% confidence intervals about the MLEs. The panel to the right shows the MLE for each harmonic, integrating over the entire data set (note the expanded vertical scale).  }
    \label{fig:mle-vs-time}
\end{figure}

\subsection{Statistical procedure}
For each harmonic $f_i$, we define the following likelihood function,
\begin{equation} \label{eq:likelihood}
\begin{alignedat}{2}
\Like_i &(\alpha, \lambda) = && \\
&\prod_{j} \bigg( \frac{1}{\sqrt{2 \pi \sigma_{ij}}}\bigg)^2 \exp && \bigg\{ \frac{-[\Re(F_{ij} - \tau_i(\alpha, \lambda, \vec{x}_j))]^2}{2 \sigma_{ij}^2} \\
& &&- \frac{[\Im(F_{ij} - \tau_i(\alpha, \lambda, \vec{x}_j))]^2}{2 \sigma_{ij}^2} \bigg\},
\end{alignedat}
\end{equation}
\noindent where $F_{ij}$ is the value of the single-sided Fourier transform of the $z$-force (normalized to units of N/$\sqrt{\rm Hz}$) in the frequency bin corresponding to $f_i$, $\tau_i(\alpha, \lambda, \vec{x}_j)$ is the value of the Fourier transform of the expected signal force in the same frequency bin for a given $\alpha$ and $\lambda$ and AS displacement $\vec{x}_j$, $\sigma_{ij}$ is the standard deviation of the Gaussian white noise in the frequency bin for $f_i$, estimated from 10 neighboring sidebands, $j$ indexes the $10^4$, 10-second-long, data-files, and $\Re()$ and $\Im()$ are the real and imaginary components of the complex-valued Fourier transforms, respectively. 

Specifically, $\sigma_{ij}$ is calculated as follows for a single harmonic, $f_i$, and continuous integration, $j$, from the observed variance of neighboring sidebands $f_k$:
\begin{equation} \label{eq:uncertainty}
\begin{aligned}
\sigma_{ij}^2 = \frac{1}{2 N_{\rm sb}} \sum_{k=1}^{N_{\rm sb}} \left[ \Re(F_{kj})^2 + \Im(F_{kj})^2 \right],
\end{aligned}
\end{equation}
\noindent where $N_{\rm sb}=10$ is the number of sidebands, $F_{kj}$ is the value of the Fourier transform of the $z$-force in the frequency bin corresponding to the sideband $f_k$, and the factor of (1/2) yields the expected uncertainty for either the real or imaginary component independently.

Each $\Like_i(\alpha,\lambda)$ can be used individually to provide the maximum likelihood estimator, $\hat{\alpha}_i$, for each harmonic, as shown in Fig.~\ref{fig:mle-vs-time}. It is confirmed that the measured signals are background-like and not due to a novel interaction by  observing that the amplitudes extracted for each selected harmonic do not exhibit the expected ratio from the signal as shown in Fig.~\ref{fig:asd}. In addition, the expected time-invariant behavior is not found in the data. 

Due to the different levels of background in different harmonics, each is treated independently in the statistical procedure and combined in an approach following ~\cite{Cowan:2010}. This utilizes the fact that a gravity-like force should be present in all harmonics, increasing the sensitivity when backgrounds are correlated differently than the expected signal.

Since the described experiment is sensitive to the direction of the force, upper limits can be set separately on positive and negative values for $\alpha$. Harmonics with $\hat{\alpha}_i>0$ are used to constrain an upper limit on $\alpha>0$, while those with $\hat{\alpha}_i<0$ constrain $\alpha<0$, following the procedure in~\cite{Cowan:2010}. A test statistic for harmonics $f_i$ with $\hat{\alpha}_i>0$ is defined as
\begin{equation} \label{eq:log-likelihood}
q_{\alpha,i} = 
    \begin{cases}
    -2 \log \left( \frac{\Like_i(\alpha,\lambda)}{\Like_i(\hat{\alpha}_i,\lambda)} \right) & \alpha \geq \hat{\alpha}_i \\
    \quad \quad 0 & \alpha < \hat{\alpha}_i \\
    \end{cases},
\end{equation}
\noindent where a nearly identical function is defined for harmonics with $\hat{\alpha}_i<0$, but with the conditions flipped appropriately for the change in sign. The final test statistic used to establish upper limits on alpha is simply the sum over all harmonics, $q_{\alpha}=\sum_i q_{\alpha,i}$, and is profiled independently for $\alpha>0$ and $\alpha<0$. For this work, the entire procedure was completed with three completely independent analysis frameworks, in order to provide a level of cross-validation.

The method introduced above was thoroughly investigated by injecting artificial software signals on top of actual experimental noise. Data sets with a total length of $10^{4}~$seconds were used, in which the relative positions of MS and AS are nearly the same as in the primary measurement, but with no scanning motion and hence with no signal or background. This was done repeatedly for a range of both parameters, and an upper limit was estimated for each unique data set. This process validates the analysis, quantifying the deviation from Wilk’s theorem~\cite{Cowan:2010} (and the expected $\chi^2$ distribution), and finding the critical values corresponding to the 95\% CL upper limit. In a separate process, constant and time-varying backgrounds were added together with a simulated signal, testing various scenarios and demonstrating that the procedure is robust against undercoverage.


\section{Results}

\begin{figure}[!t]
    \includegraphics[width=1\columnwidth]{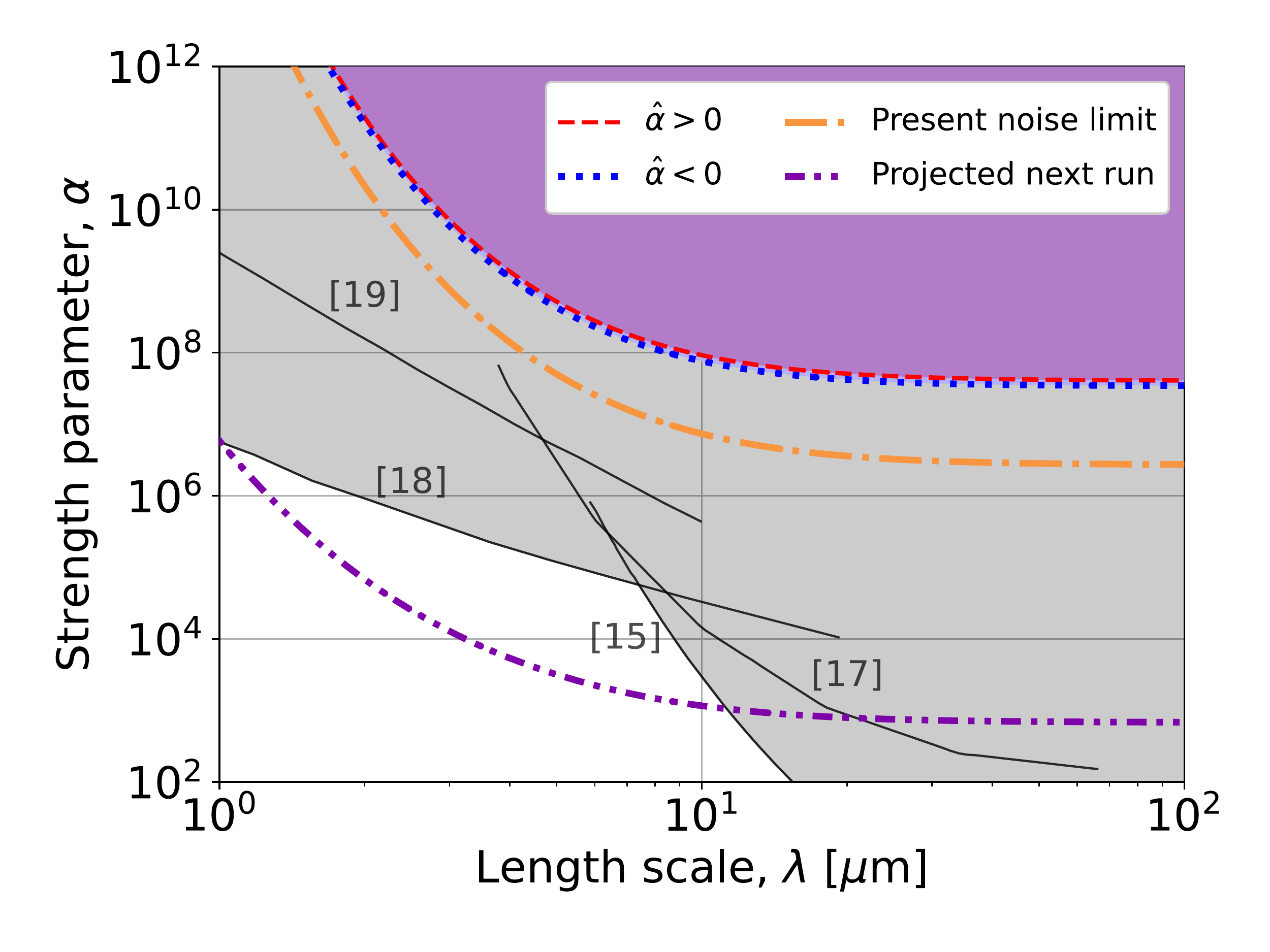}
    \caption{Limit curve in the $\alpha - \lambda$ parameter space. The region above and to the right of the red and blue lines indicates the parameter space excluded by this experiment for positive and negative $\alpha$, respectively, with a 95\% confidence level. The gray region shows the parameter space covered by previous searches~\cite{Geraci:2008,Sushkov:2011,Chen:2016,Lee:2020}. The background-free sensitivity for this run, using the current AS-MS separation, noise conditions, and integration time is shown by dash-dotted orange line. In addition, the projected sensitivity for the next run, given the improvements outlined in the text, is shown by the dash-double-dotted purple line.  This assumes a noise floor of $1\times10^{-18}~{\rm N}/\sqrt{\rm Hz}$, AS-MS separation of 7.5 (-5) $\mu$m in the $x$ ($z$) direction, and an integration time of 30 days. The reach could be extended further by using larger microspheres~\cite{Monteiro:2020} or smaller separations~\cite{Kawasaki:2020}.}
    \label{fig:limit}
\end{figure}

For values ranging from $\lambda=1~\mu$m to $\lambda=100~\mu$m the results are shown in Fig.~\ref{fig:limit}. The proximity of the upper limit on $|\alpha|$ for both directions implies that the background is of the same order of magnitude in the most sensitive harmonics. This provides a degree of robustness against possible cancellations with backgrounds and signal in opposite directions, as the expected signals (in terms of $\hat{\alpha}$) should be consistent between harmonics. The limit is constant for $\lambda\gtrsim10~\mu$m, and degrades exponentially as the length scale becomes shorter than the separation between the AS and the MS.

\begin{table}[!b]
\begin{center}
\caption{List of systematic uncertainties.}
\label{tab:sysuncert}
\begin{tabular}{c c c}
    \hline
    \hline
        \textbf{Effect $\epsilon$} & \textbf{$\Delta \epsilon$}& \textbf{$\Delta\alpha / \alpha$}  \\
    \hline
        Drift of amplitude response & $10\%$ & $10\%$ \\
        Attractor thickness & $1~\mu$m & $11\%$ \\
        Phase response & ${\sim}0.1~$rad & $12\%$ \\
        Distances in Y & ${<}0.2~\mu$m & $<3\%$ \\
        Distances in Z & ${<}0.9~\mu$m & $<6\%$ \\
        Distances in X & $1.5~\mu$m & $30\%$ \\
        MS weight & $15~$pg & $3.5\%$  \\
    \hline
    \hline
\end{tabular}
\end{center}
\end{table}

The main systematic uncertainties are summarized in Table~\ref{tab:sysuncert}. The dominant effect is the uncertainty in the distance between the AS and the MS in the $x$ direction. Further significant contributions come from uncertainties in the phase response of the MS as measured in the calibration procedure, uncertainty about the AS thickness, as well as drift of the amplitude response. MS properties, distances in $x$ and $y$, and alignment stability and accuracy of the AS movement have been found negligible.

The main limitation of the investigation presented here is the existence of backgrounds originating from electrostatic interactions, stray light modulated by the AS motion, and vibrations of components inside the vacuum chamber. As mentioned above, the interaction between the MS and an electric field gradient arising from a contact potential can be calculated to provide an adequate model for the electrostatic backgrounds. This model can be constrained and validated by a three-dimensional scan in which the AS is placed in different regions around the MS, as in Ref.~\cite{Blakemore:2019}. The interaction will then be minimized by applying a bias between shield and AS to null the contact potential. The stray light background is being investigated with a combination of measurements and ray tracing analysis, to inform the design of light baffles, in parallel with the development of a new multi-pixel sensor to replace the QPD, that will provide discrimination between actual shifts of the MS and changes in the halo.  Finally, critical optical components inside the vacuum chamber are being stiffened to minimize vibrations.  Those efforts, along with the multi-harmonic analysis technique presented above, are expected to push the experiment into the noise dominated regime for the next run.  The improvement in sensitivity from those changes, without altering other parameters, can be seen in Fig.~\ref{fig:limit}.

Beyond background suppression, an improvement in sensitivity in terms of noise reduction is targeted with the next iteration of the experiment. It is important to emphasize that the force sensitivity of the system is limited by pointing fluctuation of the trap beam~\cite{Kawasaki:2020} and not by shot noise or residual gas damping as already demonstrated in Refs.~\cite{Ranjit:2016,Monteiro:2020}. Therefore, an enclosure of the input external to the vacuum chamber, possibly replacing air with helium to lower the refractive index and hence the effects from its fluctuation, along with the stiffening of mechanical components, are expected to lead to substantial reduction of the noise floor, down to ${\sim}1\times10^{-18}~{\rm N}/\sqrt{\rm Hz}$ as demonstrated in Ref.~\cite{Monteiro:2020}. In addition, a significant gain in sensitivity will be achieved by changing the position and proximity of the AS, which, in the current run was limited by misalignment of the AS and the electrostatic shield. The projected sensitivity assuming the lower noise floor, 7.5 (-5) $\mu$m separation in the $x$ ($z$) direction, and an integration time of 30~days is shown in Fig.~\ref{fig:limit}.

\section{Conclusion}

We have described the results of the first experiment searching for non-Newtonian forces which couple to mass using optically levitated test masses. The effects observed in the data are not consistent with a new interaction, and the result is interpreted in terms of upper limits on the Yukawa parameter $\alpha$. These are $\alpha>9\times10^7$ and $\alpha<-8\times10^7$ with 95\% confidence level at $\lambda=10~\mu$m. The length scales involved in the experiment, in terms of dimensions of the test masses and feature size of the source of the interaction, and the separation between the two, are, for the first time, all similar to the characteristic length scale being probed. Therefore, this method provides a more robust test that applies also for interactions that cannot be parameterized with a Yukawa potential.  Substantial improvements in sensitivity are expected for the next round of measurements.

\section{acknowledgments}
This work was supported, in part, by NSF grant PHY1802952, ONR grant N00014-18-1-2409, and the Heising-Simons Foundation.  Fabrication and characterization of both the attractor and shield were performed in the nano@Stanford labs and Stanford Nano Shared Facilities (SNSF), both of which are supported by the National Science Foundation as part of the National Nanotechnology Coordinated Infrastructure under Award No. ECCS-1542152. C.P.B. acknowledges the partial support of a Gerald~J. Lieberman Fellowship of Stanford University. A.K. acknowledges the partial support of a William~M. and Jane~D. Fairbank Postdoctoral Fellowship of Stanford University. N.P. acknowledges the partial support of the Koret Foundation. We acknowledge regular discussions on the physics of trapped microspheres with the group of Prof. D.~Moore at Yale.  We also thank M.~Lu, S.~Roy, who contributed to early developments of the experimental apparatus.  Finally, we thank the personnel of the Physics machine shop at Stanford for their skilled mechanical support.

\bibliography{modified_gravity}

\end{document}